\documentclass[fleqn,usenatbib]{mnras}
\usepackage{newtxtext,newtxmath}

\usepackage[T1]{fontenc}
\DeclareRobustCommand{\VAN}[3]{#2}
\let\VANthebibliography\thebibliography
\def\thebibliography{\DeclareRobustCommand{\VAN}[3]{##3}\VANthebibliography}

\usepackage{graphicx}	
\usepackage{amsmath}	
\usepackage{subcaption}  
\usepackage{caption}
\usepackage[normalem]{ulem}
\usepackage{xcolor}
\definecolor{mediumorchid}{rgb}{0.73, 0.33, 0.83}
\usepackage{array,multirow,graphicx}
\usepackage{bigints}
\usepackage{fontawesome5}
\usepackage{hyperref}
\usepackage{graphicx}	
\usepackage{amsmath}	
\usepackage{adjustbox}

\newcommand{\github}[1]{%
   $\,$\href{#1}{\textcolor{mediumorchid}{\large\faGithubSquare}}}

\newcommand{\be}{\begin{equation}}
\newcommand{\ee}{\end{equation}}
\newcommand{\bea}{\begin{eqnarray}}
\newcommand{\eea}{\end{eqnarray}}

\title[IDE forecasts through differential emulators]{Testing interacting dark energy with Stage IV cosmic shear surveys through differentiable neural emulators}

\author[K. Carrion et al.]{
Karim Carrion,$^{1,2}$\thanks{E-mail: \href{mailto:kcarrion@icf.unam.mx}{kcarrion@icf.unam.mx}}
Alessio Spurio Mancini,$^{2}$
Davide Piras$^{3}$
and Juan Carlos Hidalgo$^{1}$\thanks{E-mail: \href{mailto:hidalgo@icf.unam.mx}{hidalgo@icf.unam.mx}}
\\
$^{1}$Instituto de Ciencias F\'{\i}sicas, Universidad Nacional Aut\'onoma de M\'exico, Cuernavaca, Morelos, 62210, Mexico\\
$^{2}$Department of Physics, Royal Holloway, University of London, Egham Hill, Egham TW20 0EX, UK\\
$^{3}$Centre Universitaire d’Informatique, Université de Genève, 7 route de Drize, 1227 Genève, Switzerland
}

\date{Accepted XXX. Received YYY; in original form ZZZ}

\pubyear{\the\year{}}

\begin{document}
\label{firstpage}
\pagerange{\pageref{firstpage}--\pageref{lastpage}}
\maketitle

\begin{abstract}
We employ a novel framework for accelerated cosmological inference, based on neural emulators and gradient-based sampling methods, to forecast constraints on dark energy models from Stage IV cosmic shear surveys. We focus on dark scattering (DS), an interacting dark energy model with pure momentum exchange in the dark sector, and train \texttt{CosmoPower} emulators to accurately and efficiently model the DS non-linear matter power spectrum produced by the halo model reaction framework, including the effects of baryon feedback and massive neutrinos.
We embed them within a fully-differentiable pipeline for gradient-based cosmological inference for which the batch likelihood call is up to $\mathcal{O}(10^5)$ times faster than with traditional approaches, producing constraints from simulated Stage IV cosmic shear data running on a single GPU. We also perform model comparison on the output chains from the inference process, employing the learnt harmonic mean estimator implemented in the software \texttt{harmonic}.
We investigate degeneracies between dark energy and systematics parameters and assess the impact of scale cuts on the final constraints. Assuming a DS model for the mock data vector, we find that a Stage IV survey cosmic shear analysis can constrain the DS amplitude parameter $A_{\mathrm{ds}}$ with an uncertainty roughly an order of magnitude smaller than current constraints from Stage III surveys, even after marginalising over baryonic feedback, intrinsic alignments and redshift distribution uncertainties. These results show great promise for constraining DS with Stage IV data; furthermore, our methodology can be straightforwardly extended to a wide range of dark energy and modified gravity models.
\end{abstract}




\section{Introduction}
\label{sec:intro}

Ongoing and forthcoming Stage IV galaxy surveys such as the Dark Energy Spectroscopic Instrument \citep[DESI,][]{DESI:2016fyo}\footnote{\url{https://www.desi.lbl.gov/}}, \textit{Euclid}\footnote{\url{https://www.euclid-ec.org/}} \citep{Euclid:2024yrr}, the Legacy Survey of Space and Time at the Vera C. Rubin Observatory\footnote{\url{https://www.lsst.org/}} \citep[LSST,][]{LSSTDarkEnergyScience:2012kar} or the Nancy Grace Roman Space Telescope\footnote{\url{https://roman.gsfc.nasa.gov/}} \citep{2015arXiv150303757S} will generate datasets of unprecedented size and quality, with the potential to strongly constrain the models of the governing physics on cosmological scales. 

Analysing these datasets will require the exploration of high-dimensional, complex parameter spaces, to ensure accurate modelling of various systematic effects, particularly on non-linear scales which often demand computationally intensive theoretical predictions \citep[e.g.][]{Carrasco:2012cv,Cataneo:2019fjp, Bose:2020wch,Aviles:2020cax}. These challenges call for the development of novel, more efficient approaches to theoretical modelling and parameter estimation. Furthermore, it is critical to develop novel methods for rigorous, quantitative model comparison \citep[see e.g.][for a review]{Trotta:2008qt}, ideally decoupling this task from that of parameter estimation, to allow for more flexibility in the posterior sampling strategy.

In this paper, we extend previous work presented in \citet{Carrion:2024itc} and \citet{Piras:2024dml} to compute forecasts on the constraining power of Stage IV surveys on dark energy, using a novel, accelerated framework for parameter estimation and model comparison. Specifically, we focus on the pure momentum exchange in interacting dark energy (IDE) theories, where the dark sector is coupled through an elastic interaction referred to as dark scattering (DS). IDE models have demonstrated their potential to reduce some of the recently reported cosmological tensions, particularly in the Hubble constant $H_0$ \citep{DiValentino:2021izs,YAO2023101165} and the matter clustering parameter $S_8$ \citep{DiValentino:2019ffd,Lucca:2021dxo,ManciniSpurio:2021jvx,Sabogal:2024yha}. While the focus of this paper is on the DS model, the methodology presented here is general and can be easily extended to a wide range of dark energy and modified gravity models.

Our forecasts consider configurations typical for a Stage IV cosmic shear survey. We run our forecasts using a fully-differentiable pipeline which allows us to use gradient-based methods for posterior sampling. The cosmic shear power spectrum is modelled including systematics contributions arising from uncertain redshift distributions and intrinsic alignments. To accurately model the non-linear clustering in the DS model, including the contribution of massive neutrinos, we employ the halo model reaction framework \citep{Cataneo:2018cic}, which we accelerate by producing neural emulators using the package \texttt{CosmoPower} \citep{SpurioMancini:2021ppk}. We also use \texttt{CosmoPower} to emulate the baryon feedback boost to the non-linear matter power spectrum produced by \texttt{HMcode} \citep{Mead:2015yca}. Finally, we perform model comparison on the inference chains using the learnt harmonic mean estimator of the Bayesian model evidence, implemented in the software \texttt{harmonic} \citep{mcewen2021, Polanska24}.

This paper is organised as follows. In Sec.~\ref{sec:DS_model}, we briefly describe the DS model. Sec.~\ref{sec:emulators} reviews the \texttt{CosmoPower} emulators employed in our analysis. In Sec.~\ref{sec:pipeline} we describe our inference pipeline, including details on cosmic shear modelling, gradient-based sampling methods and model comparison through \texttt{harmonic}. We present our results in Sec.~\ref{sec:forecast} and draw our conclusions in Sec.~\ref{sec:conclusion}.

\section{Dark scattering}
\label{sec:DS_model}

Interacting dark energy \citep[IDE,][]{Pourtsidou:2013nha,Skordis:2015yra} theories incorporate a (model-dependent) coupling $J_\nu$ into the dark sector, such that their individual energy-momentum tensors $T^{\mu \nu}$ are no longer conserved: 
\bea
\nabla_\mu T^{\mu \nu}_{\rm DM} = J^\nu \quad \Longleftrightarrow \quad \nabla_\mu T^{\mu \nu}_{\rm DE} = -J^\nu.
\label{eq:IDE_coupled}
\eea
In these scenarios dark energy (DE) and dark matter (DM) exchange energy and/or momentum depending on the characteristics of the coupling. Such couplings can be derived from three different general IDE models (see \citealp{Pourtsidou:2013nha} for details). In particular, our model of interest involves a coupling that only depends on the fluid velocity $\mathbf{v}$ of the dark sector, known as dark scattering \citep[DS,][]{Simpson:2010}. The pure momentum exchange coupling \footnote{The DS model assumes there is no background energy exchange, leaving the background evolving as in the uncoupled case.} is given by
\bea
 \mathbf{J}_{{\rm DS}}=  -(1+w)\, \xi \, \rho_{\rm c} \, \rho_{\rm DE} \, (\mathbf{v}_{{\rm c}}-\mathbf{v}_{{\rm DE}}) \, ,
 \label{eq:DS_eq}
\eea
where $\xi$ is the coupling strength parameter (with units $\rm b/GeV$), $\rho_{\rm c}$ is the density of cold dark matter (CDM) particles and $\rho_{\rm DE}$ the dark energy density with an equation of state parameter $w$.  {The cross-section is effectively between CDM particles and DE, and is defined as $\sigma_{\rm DS} \equiv \xi \, m_{\rm c}$, where $m_{\rm c}$ is the mass of CDM particle.}

\begin{figure}
\centering 
\includegraphics[width=.5\textwidth]{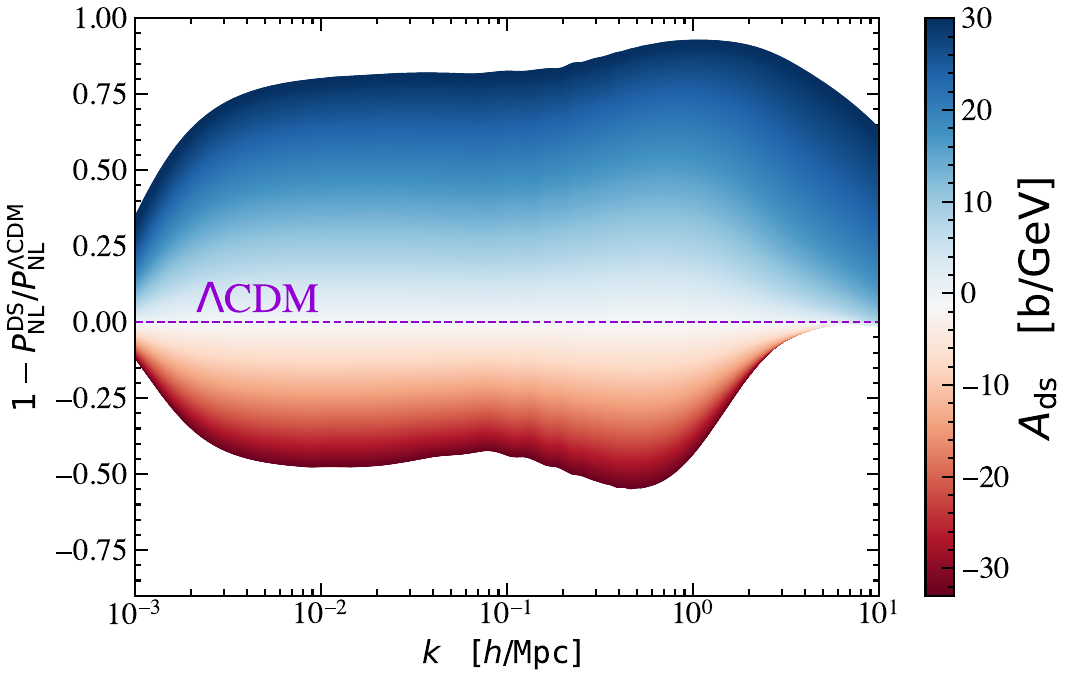}
\caption{Deviation of the non-linear matter power spectrum from that of a $\Lambda$CDM model (represented in purple dashed-line) in the presence of DS interaction at $z=0$. The color code on the right indicates that values of $A_{\rm ds}<0$ (warmer colors) enhance the matter power spectrum, while $A_{\rm ds}>0$ values (colder colors) suppress it.  {The DS emulator accurately recovers the $\Lambda$CDM case, allowing a direct comparison of deviations induced by DS interaction.}}
\label{fig:spectra_impact}
\end{figure}

The DS non-linear matter power spectrum has been validated through analytical halo modelling \citep{Carrilho:2021rqo} in comparison with full $N$-body simulations \citep{Baldi:2014, Baldi:2016zom}. If massive neutrinos are considered, then an effective strength coupling $\bar\xi$ would be introduced and modulated by the dark matter fraction $f_{\rm c}=\rho_{\rm c}/\rho_{\rm m}$ of the total matter (see \citealp{Carrilho:2021rqo} for details) as
\bea
\bar\xi=\frac{f_{\rm c}}{1+(1-f_{\rm c})\Xi(0)}\xi\,.
\label{eff_xi}
\eea
Here we have defined the evolving interaction term as
\bea
\Xi(z) \equiv A_{\rm ds} \dfrac{3 \Omega_{\rm DE}}{8 \pi G} H \, ,
\label{eq:Interaction_term}
\eea
where the effective interaction amplitude, $A_{\rm ds}$, is given by
\bea
A_{\rm ds} \equiv \bar\xi \left(1+w\right). 
\label{eq:Interaction_term_eff}
\eea
In Eq.~(\ref{eq:Interaction_term}$), H=\dot a/a$ is the Hubble rate for the scale factor $a$. Remarkably, Eq.~(\ref{eq:Interaction_term}) depends only on background quantities. Although the model is applicable to a general evolution of the equation of state parameter $w(z)$, we will take it to be constant in this work. It is clear in Eq.~(\ref{eq:Interaction_term}) that when $w \rightarrow -1$ we recover the $\Lambda$CDM model. The effects of DS and massive neutrinos are included in the equation of motion of particles as 
\bea
\dot{\mathbf{u}}_{\rm c} = - (1+\Xi) H \mathbf{u}_{\rm c} - \nabla_\mathbf{r} \Phi,
\label{eq:interaction_eom}
\eea
where $\mathbf{u}_{\rm c}$ is the comoving particle velocity and $\Phi$ is the gravitational potential.
In Eq.~(\ref{eq:interaction_eom}) the effect of the interaction manifests as an additional frictional force, dragging or pushing over the CDM particles. As expected, this alters the matter power spectrum depending on ${\rm{sign}}(A_{\rm ds})$. Fig.~\ref{fig:spectra_impact} illustrates the influence of the interaction on the non-linear matter power spectrum.

\section{\texttt{CosmoPower} emulators}
\label{sec:emulators}

We train two emulators using \href{https://github.com/alessiospuriomancini/cosmopower}{\texttt{CosmoPower}} \citep{SpurioMancini:2021ppk}: the first one emulates the DS non-linear matter power spectrum, while the second one provides the baryonic correction on the matter power spectrum. Both emulators are publicly available at \href{{https://github.com/karimpsi22/DS-emulators}}{\texttt{DS-emulators}}.

\subsection{Dark scattering matter power spectrum emulator}
\label{subsec:reaction}

Our main emulator of the DS non-linear power spectrum is based on the halo model reaction framework (see \citealp{Carrilho:2021rqo} for validation against $N$-body simulations). This formalism for the halo model is encoded in a function called reaction, $\mathcal{R}(k,z)$, related to the non-linear matter power spectrum as
\bea
 P^{\rm alt}_{\rm NL}(k,z) \equiv \mathcal{R}(k,z) \times   P_{\rm NL}^{\rm pseudo}(k,z) \, .
 \label{eq:Reaction_spectrum}
\eea
The reaction function may account for modifications of the alternative theory (denoted by the superscript `alt'), such as massive neutrino contributions or a non-standard dark sector or modified gravity. Its full expression is given in the following form:
\bea
    \mathcal{R}(k,z)=\frac{\bar{f}_{\nu}^{2} P_{\mathrm{HM}}^{(\mathrm{cb})}(k,z)+2 f_{\nu}\bar{f}_{\nu} P_{\mathrm{HM}}^{(\mathrm{cb} \nu)}(k,z)+f_{\nu}^{2} P_{\mathrm{L}}^{(\nu)}(k,z)}{P_{\mathrm{L}}^{(\mathrm{m})}(k,z)+P_{\mathrm{1h}}^{\mathrm{pseudo}}(k,z)}  \, .
    \label{eq:reaction_def}
\eea
The reaction enables the connection between the spectrum of the alternative theory to a quasi-$\Lambda$CDM cosmology spectrum (denoted by the superscript `pseudo'). The superscripts `L' refer to linear and `1h' to 1-halo term. The `pseudo' cosmology serves as condition of matching, specifically, both linear power spectrum $P^{\rm alt}_{\mathrm{L}}(k,z) = P^{\rm pseudo}_{\mathrm{L}}(k,z)$ are matched at the desired redshift $z$. In Eq.~(\ref{eq:reaction_def}) the superscript $\rm (m) \equiv (cb+\nu) $ accounts for the sum of matter components, where the sum of CDM and baryons is represented by the superscript `cb', and `$\nu$' stands for massive neutrinos; finally, $\bar{f}_{\nu} = 1 - f_{\nu}$, where  $f_\nu = \rho_\nu/ \rho_\mathrm{m}$ is the ratio between massive neutrinos and total matter densities.  {We assume 1 massive and 2 massless neutrinos throughout this paper.}

We use the same emulator trained in \citet{Carrion:2024itc}, and refer to that paper for details on the emulator training and validation in the region $-3 < \log_{10} (k\, \text{Mpc}) < 1$ and $0 < z < 5$. The input of the emulator is given by the following set of parameters: 
\bea 
\boldsymbol{\theta}_{\rm DS} = \{\omega_{\rm b}, \omega_{\rm cdm}, h, n_{\rm{s}}, S_8, m_\nu, w, A_{\rm ds}, z \},
\label{eq:DS_parameters_training}
\eea
where $\omega_i = \Omega_i h^2$ is the physical density parameter for $i= \{\rm b, cdm\}$, $h$ is the reduced Hubble parameter, $n_{\rm{s}}$ is the scalar spectral index and $m_\nu$ the sum of neutrino masses. The accuracy achieved is sub-percent with respect to the predictions of \texttt{ReACT} (see \citealp{Carrion:2024itc} for details).
The halo model reaction is publicly available in the latest version of the \href{{https://github.com/nebblu/ACTio-ReACTio}}{\texttt{ReACT}} code;  {it takes the linear matter power spectrum as an input, which has been obtained from a modified version of \href{{https://github.com/PedroCarrilho/class_public/tree/IDE_DS}}{\texttt{CLASS}}.} We refer the reader to \cite{Cataneo:2018cic,Bose:2020wch,Bose:2021mkz} for further details on the modelling.

\begin{table}
  \centering 
   \caption{Overview of the prior distributions applied to cosmological and nuisance parameters in the simulated analyses. The specified ranges for cosmological and baryonic feedback parameters ensure compatibility with the validity limits of the emulators used in this work. The last column presents the fiducial values of the cosmological parameters, where the values for the $\Lambda$CDM case are taken from the best-fit results of Planck 2018 TTEETE+BAO \citep{Planck:2018vyg}. In the DS model we allow for a non-zero dark sector interaction. In particular, the choice of $(w, A_{\rm ds})$ values are motivated by the best-fit findings from our previous work \citep{Carrion:2024itc}. The units of $m_\nu$ are eV, while $A_{\rm ds}$ has units of b/GeV.  {Each $D_{z_{i, \rm source}} = D_{z_{i}}$ quantifies the photometric redshift uncertainty, shifting each $i$-bin distribution as $n_i'(z) = n_i(z - D_{z_i})$.}}
\label{tab:priors_and_fiducial}  \renewcommand{\arraystretch}{1.75}
  \setlength{\tabcolsep}{2.pt}
  \begin{tabular}{c c c c}
    &\textbf{Input Parameter}  &  \textbf{Prior} &  $\Lambda$CDM / DS fiducial \\
    \hline
    \hline
     \parbox[c]{5mm}{\multirow{8.5}{*}{\rotatebox[origin=c]{90}{\textbf{Cosmology}}}} 
     &$\omega_{\mathrm{b}} = \Omega_{\mathrm{b}} h^2$ & 
    $\mathcal{U}(0.01875, 0.02625)$ 
    & $0.02242$ \, / \, $0.02242$ \\
    &$\omega_{\mathrm{cdm}} = \Omega_{\mathrm{cdm}} h^2$ & $\mathcal{U}(0.05, 0.255)$   
    & $0.11933$ \, / \, $0.11933$ \\
    &$h$ & 
    $\mathcal{U}(0.64, 0.82)$  
    & $0.682$ \, / \, $0.682$\\
    &$n_{\mathrm{s}}$ & 
    $\mathcal{U}(0.84, 1.1)$  
    & $0.9665$ \, / \, $0.9665$\\
    &$S_8$ &  
    $\mathcal{U}(0.6, 0.9)$  
    & $0.825$ \, / \, $0.825$ \\
    &$m_\nu$ &  
    $\mathcal{U}(0,0.2)$  
    & $0.06$ \, / \, $0.06$ \\
    &$w$  & 
    $\mathcal{U}(-1.3, -0.7)$   
    & $-1$ \, / \, $-0.967$\\ 
    &$|A_{\mathrm{ds}}|$ & 
    $\mathcal{U}(0, 30)$  
    & $0$ \, / \, $10.6$ \\[1.ex]
    \hline
    \hline
    \\[-3.5ex] \parbox[c]{5mm}{\multirow{1.5}{*}{\rotatebox[origin=c]{90}{\textbf{Baryons}}}}
     &$c_{\mathrm{min}}$  & 
     $\mathcal{U}(2, 4)$   
     & $2.6$ \, / \, $2.6$  \\[3.ex]
    \hline
    \hline
    \\[-5ex] \parbox[c]{5mm}{\multirow{2.5}{*}{\rotatebox[origin=c]{90}{\textbf{IA}}}} 
     &$A_{\mathrm{IA}}$  & 
     $\mathcal{U}(-5, 5)$  
     & $0.8$ \, / \, $0.8$ \\
    &$\eta_{\mathrm{IA}}$   & 
    $\mathcal{U}(-5, 5)$   
    & $0$ \, / \, $0$ \\[1ex]
    \hline
    \hline
    \\[-3.5ex] \parbox[c]{5mm}{\multirow{1.5}{*}{\rotatebox[origin=c]{90}{\textbf{$z$-bins}}}} 
     & $D_{z_i, \rm source}$ \, $i = 1, \cdots 10$ &            $\mathcal{N}(0, 10^{-4})$ 
     & $0$ \, / \, $0$  \\[2.5ex]
    \hline
    \hline
    \end{tabular}
\end{table}

\subsection{Baryonic feedback}

The effect of baryonic feedback \citep[see e.g.][for a review]{Chisari:2019tus} on the full power spectrum $\hat{P}(k,z)$ is taken into account as a boost $\mathcal{B}(k,z)$ \citep{Arico_2021, Giri:2021qin} to the DM-only matter power spectrum:
\bea
\mathcal{B}(k,z) = \dfrac{\hat{P}(k,z)}{P_{\rm DM\text{-}only}(k,z)} \, .
\label{eq:boost_eq}
\eea
We calculate the non-linear power spectrum by multiplying the emulated components:
\bea
\hat{P}_{\rm NL}(k,z) = \mathcal{B}(k,z) \times P^{\rm alt}_{\rm NL} (k,z) \, .
\label{full:spectrum}
\eea
To compute Eq.~(\ref{eq:boost_eq}) we emulate the \texttt{HMcode2016} \citep{Mead:2015yca} prescription, which depends on two baryonic parameters $c_{\rm min}$ and $\eta_0$ capturing the influence of baryons within a halo. Here, $\eta_0$ is determined by the formula $\eta_0 = 0.98 - 0.12 \, c_{\rm min}$, obtained in \citet{Mead:2015yca} by fitting to the OverWhelmingly Large (OWL) hydrodynamical simulations \citep{2010MNRAS.402.1536S,2011MNRAS.415.3649V,2011MNRAS.417.2020S}. 

The \texttt{CosmoPower} emulator for the \texttt{HMcode2016} baryonic feedback prescription has the following input parameters:
\bea
\boldsymbol{\theta}_{\rm feedback} = \{\omega_{\rm b}, \omega_{\rm cdm}, h, n_s, S_8, c_{\rm min}, \eta_0 ,z \}.
\label{eq:boost_parameters_training}
\eea
We train this emulator in the same ranges of $k$-mode and redshift used for the \texttt{ReACT} emulator. 

\section{Inference pipeline}
\label{sec:pipeline}

\subsection{\texttt{JAX} framework}

We use \href{{https://github.com/dpiras/cosmopower-jax}}{\texttt{CosmoPower-JAX}} \citep{Piras:2023aub} to load and manage our emulators. \texttt{CosmoPower-JAX} is a flexible implementation of \texttt{CosmoPower} in \href{{https://github.com/google/jax.git}}{\texttt{JAX}} \citep{jax2018github}. 
\texttt{JAX} allows for efficient automatic differentiation of functions written in \texttt{Python}, and enables the dynamic compilation and batch evaluation of \texttt{NumPy} \citep{harris2020array} operations, allowing them to run efficiently on graphics processing units (GPUs) and tensor processing units (TPUs). In this work, we employ these \texttt{JAX} features to significantly enhance the efficiency of Bayesian cosmological inference, obtaining orders-of-magnitude acceleration.

\subsection{Cosmic shear power spectrum modelling}
\label{subsec:shear_model}

Throughout our pipeline, we employ the \texttt{JAX}-based library \href{{https://github.com/DifferentiableUniverseInitiative/jax_cosmo}}{\texttt{jax-cosmo}} \citep{Campagne:2023ter} to model the cosmic shear power spectrum $\mathcal{C}_{i j}^{\gamma \gamma}(\ell)$ as a function of the multipole $\ell$. Given $N_{\rm bins}$ tomographic redshift bins, the cosmic shear power spectrum for a pair of bins $i, j = 1, \dots , N_{\rm bins}$ is an integral of the non-linear matter power spectrum $\hat{P}_{\rm NL}(k,a)$ along the line of sight:
\bea
\mathcal{C}_{i j}^{\gamma \gamma}(\ell) = {\bigintsss\limits_{0}}^{\chi_{H}} \mathrm{d}\chi \, \frac{W_i^\gamma \, W_j^\gamma }{\chi^2} \, \hat{P}_{\rm NL} \left(k = \frac{\ell + 1/2}{\chi}, a \right),
\label{eq:cell_shear}
\eea
where $\chi$ is the comoving distance  (depending itself on the scale factor $a$), and the distance at the upper limit of the integral defines the Hubble radius $\chi_{H} = c/H_0$. Note that in Eq.~(\ref{eq:cell_shear}) the Fourier $k$-modes are linked to the angular multipoles $\ell$ through the extended Limber approximation\footnote{The accuracy of the extended Limber approximation was not explored in this paper.} \citep{LoVerde:2008re}. Given a distribution $n_{s,i}(z)$ for each redshift bin $i = 1, \dots , N_{\rm bins}$, and assuming spatial flatness, the weighting function $W_i^\gamma$ is computed as
\bea
    W_i^\gamma (\chi) = \dfrac{3}{2} \Omega_{\rm m} \dfrac{H_0^2}{c^2} \dfrac{\chi}{a} {\bigintsss\limits_{\chi}}^{\chi_{\rm H}} \mathrm{d} \chi' \, n_{s,i}(\chi') \, \dfrac{\chi'-\chi}{\chi'} ,
\label{eq:window_shear}
\eea
where $\Omega_{\rm m}$ is the total matter density parameter and $H_0$ is the Hubble constant. Eq.~(\ref{eq:cell_shear}) quantifies the angular power spectrum of the total matter distribution, accounting for both dark and visible matter.

\subsection{Intrinsic alignment contribution}

The lensing signal is contaminated by the intrinsic alignment (IA) of galaxies. To address this, we add IA terms to Eq.~(\ref{eq:cell_shear}) as:
\bea
\mathcal{C}_{ij}^{\hat{\gamma} \hat{\gamma}}(\ell) = \mathcal{C}_{ij}^{\gamma \gamma}(\ell) + \mathcal{C}_{ij}^{\gamma \mathrm{I}}(\ell) + \mathcal{C}_{ij}^{\mathrm{I}  \gamma}(\ell) + \mathcal{C}_{ij}^{\mathrm{I} \mathrm{I}}(\ell) \ .
\label{eq:cell_shear_full}
\eea
The window function for the IA contribution is obtained from the commonly-used non-linear alignment (NLA) model \citep{Hirata:2004gc, Bridle:2007ft}, which is modulated by two free parameters, an amplitude $A_{\rm IA}$ and a power law parameter $\eta_{\rm IA}$:
\bea
    W_i^{\rm I}(\chi) = - A_{\rm IA} \left( \frac{1+z}{1+z_{\rm p}} \right)^{\eta_{\rm IA}} \, n_{i}(\chi) \, \frac{C_1 \, \rho_{\rm crit} \, \Omega_{\rm m}}{D_+(\chi)},
    \label{eq:window_NLA}
\eea
where $D_+(\chi)$ represents the growth factor and $\rho_{\rm crit}$ is the critical matter density. The constants $C_1$ and the pivot redshift $z_{\rm p}$ are set to $0.64$ and $0.3$, respectively.

We incorporate the \texttt{CosmoPower} emulators for DS and baryonic non-linear power spectrum into \texttt{jax-cosmo} after loading them using \texttt{CosmoPower-JAX}, replacing the \texttt{HaloFit} \citep{Takahashi:2012em} prescription for computing the non-linear power spectrum in \texttt{jax-cosmo}. 

\subsection{Inference method}
\label{subsec:inference}

To sample the posterior distribution, we employ the \texttt{NUTS} \citep[No-U-Turn Sampler,][]{hoffman2014no} algorithm as implemented in the \href{{https://github.com/pyro-ppl/numpyro}}{\texttt{NumPyro}} \citep{phan2019composable} library. This sampler is an adaptive variant of the Hamiltonian Monte Carlo (HMC) algorithm, allowing for efficient and scalable sampling in high-dimensional spaces. \texttt{NumPyro} offers compatibility with \texttt{JAX} (and therefore \texttt{jax-cosmo} and \texttt{CosmoPower-JAX}), enabling us to build a fully-differentiable inference pipeline.

The gradients for standard HMC algorithms are typically calculated using finite difference methods, which often lack precision and can be computationally expensive. In the \texttt{NUTS} framework, these inefficiencies are mitigated through adapting the sampling path to prevent it from looping back to regions of the parameter space that have already been sampled, hence optimizing the overall efficiency of the sampling process. In our case, with a differentiable pipeline in place, gradients can be efficiently computed using automatic differentiation. Thus, the key requirement is that the likelihood must be differentiable, which we achieve using \texttt{CosmoPower-JAX} and \texttt{jax-cosmo}.

This differentiable framework allows us to perform faster batch evaluations of the likelihood, taking 0.19 s on a single GPU versus $\sim 8$ h on a single CPU to produce $1\,000$ spectra. This results in an acceleration of up to $\mathcal{O}(10^5)$ with respect to traditional non-differentiable methods. This significant efficiency not only accelerates parameter inference but also facilitates the exploration of various scenarios based on Stage IV cosmic shear mock data.

\subsection{Model comparison}
\label{sec:model_comparison}

A standard method for comparing cosmological models is computing their Bayesian model evidence, which provides a quantitative approach to determine which model is favoured in light of the observed data, accounting for both model complexity and quality of the fit. In this section, we review key concepts of Bayesian model comparison \citep{Trotta:2008qt}.

Given some parameters $\boldsymbol{\theta}$, associated with an assumed model $\mathcal{M}$, and observed data $\mathcal{D}$, the Bayesian model evidence $\mathcal{Z}_\mathcal{M}$ is defined as:
\bea	
\mathcal{Z}_\mathcal{M} \equiv \mathcal{P}  ( \mathcal{D} \vert \mathcal{M} ) = \int \mathrm{d} \boldsymbol{\theta} \, \mathcal{P}  ( \mathcal{D} \vert \boldsymbol{\theta}, \mathcal{M} ) \times \mathcal{P}  ( \boldsymbol{\mathrm{\theta}} \vert \mathcal{M}),
\label{eq:evidence}
\eea
which is the probability of the data given a certain model, expressed as a function of the likelihood $\mathcal{P}  ( \mathcal{D} \vert \boldsymbol{\theta}, \mathcal{M} )$ and the prior $\mathcal{P}  ( \boldsymbol{\mathrm{\theta}} \vert \mathcal{M})$. In the context of Bayesian statistics, the evidence is derived directly from Bayes' theorem, which describes how the posterior distribution of model parameters relates to the observed data for a given model:
\bea
\mathcal{P} ( \boldsymbol{\theta} \vert \mathcal{D}, \mathcal{M} )
= \frac{\mathcal{P} ( \mathcal{D} \vert \boldsymbol{\theta}, \mathcal{M} ) \times \mathcal{P} ( \boldsymbol{\mathrm{\theta}} \vert \mathcal{M} )}{\mathcal{P} ( \mathcal{D} \vert \mathcal{M} )}.
\label{eq:bayes_def}
\eea
The term on the left-hand side represents the posterior distribution of the parameters, $\mathcal{P}(\boldsymbol{\theta} \vert \mathcal{D}, \mathcal{M})$, which is updated based on the observed data and the assumed model. The evidence, as defined in Eq.~(\ref{eq:evidence}), acts as a normalisation factor for the posterior distribution.

When performing parameter estimation for a single model, the evidence is typically ignored, while it becomes essential when multiple models are being considered. The key quantity is the evidence ratio between two competing models $\mathcal{M}_1$ and $\mathcal{M}_2$:
\bea
B = \frac{\mathcal{Z}_{\mathcal{M}_1}}{\mathcal{Z}_{\mathcal{M}_2}} =  \frac{\mathcal{P} ( \mathcal{D}  \vert \mathcal{M}_1 )}{ \mathcal{P} ( \mathcal{D} \vert \mathcal{M}_2 )}.
\label{eq:bayes_factor}
\eea
known as the Bayes factor $B$. Using Jeffreys’ scale \citep{jeffreys1998theory,Nesseris:2012cq} it is possible to use this ratio to provide a measure of how strongly the data supports $\mathcal{M}_1$ over $\mathcal{M}_2$.

\subsection{Learned harmonic mean estimator}
\label{subsec:harmonic}

Computing the evidence involves solving a complicated integral, as shown in Eq.~(\ref{eq:evidence}), which hinders its estimation especially in high-dimensional cases. There are several algorithms to compute the evidence, as part of the inference or using only samples from the posterior distribution \citep{Skilling:2006gxv,Feroz:2008xx,Buchner:2014nha,Handley:2015fda,2017MCevidence,2019arXiv191206073J,albert2020jaxns,2024arXiv240412294S}. 

In our analysis, we opt to use the \texttt{harmonic} package, which implements the learnt harmonic mean estimator with normalising flows \citep{mcewen2021, Polanska24} to estimate the Bayesian evidence. \texttt{harmonic} is agnostic to the sampling scheme and has been demonstrated to achieve accuracy comparable to other methods like nested sampling \citep{Piras:2024dml}, thus offering greater scalability using only samples from the posterior distribution and their associated probability densities.

The method trains a target distribution $\varphi(\boldsymbol{\theta})$ using a normalising flow. Once the flow can approximate the posterior while having tighter tails (imposed through a temperature parameter $T\in [0,1]$), it is used to compute the harmonic mean estimator \citep{Newton94}:
\bea
\hat{\rho} = \frac{1}{N} \sum_{i=1}^{N} \frac{\varphi (\boldsymbol{\theta}_i)}{\mathcal{P}(\mathcal{D} \vert \boldsymbol{\theta}_i ) \mathcal{P} (\boldsymbol{\theta}_i)} = \frac{1}{\mathcal{Z}}, \quad \boldsymbol{\theta}_i \sim \mathcal{P}(\boldsymbol{\theta} \vert \mathcal{D}),
\label{eq:harmonic_estimator}
\eea
where $N$ is the number of posterior samples $\{\boldsymbol{\theta}_i\}_{i=1}^N$. We refer the reader to \citet{mcewen2021, Polanska24} for more details. To estimate the evidence of each model, we apply the \texttt{RQSplineModel} \citep{Durkan:2019nsq} provided within \texttt{harmonic}, which comprises $4$ layers and $128$ spline bins. In the flow training process, 50\% of the chains are used as training samples, while the remaining chains are employed to compute the evidence; we set $T=0.8$.

\begin{figure}
\centering 
\includegraphics[width=.48\textwidth]{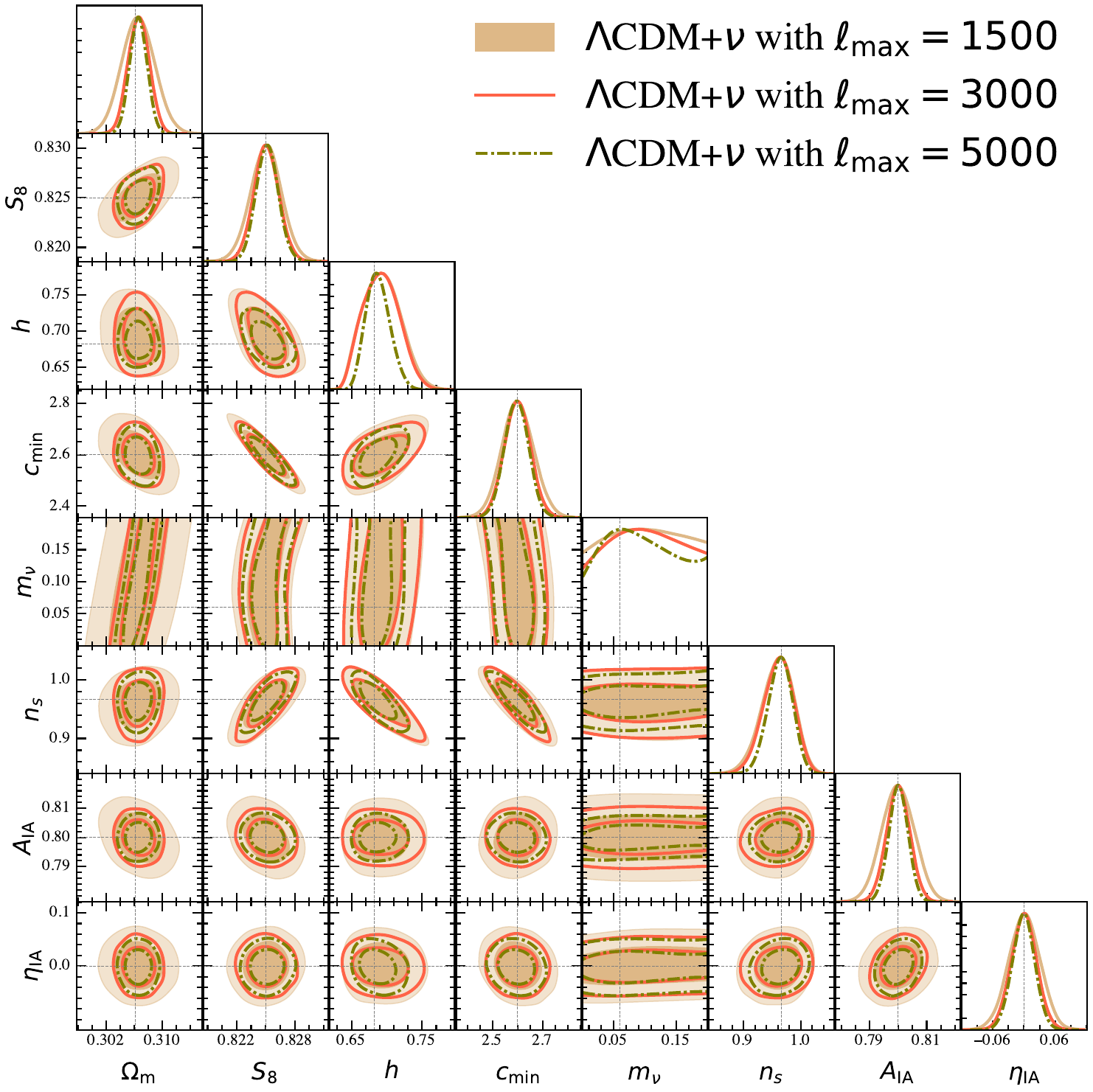}
\caption{68\% and 95\% 2D and 1D marginalised posterior contours for key cosmological and nuisance parameters for the $\Lambda$CDM model using three different scale-cuts. The thin dashed-grey straight lines indicate the $\Lambda$CDM fiducial values used to generate the mock data vector.}
\label{fig:lcdm_comparison} 
\end{figure}

\begin{figure}
\centering 
\includegraphics[width=.48\textwidth]{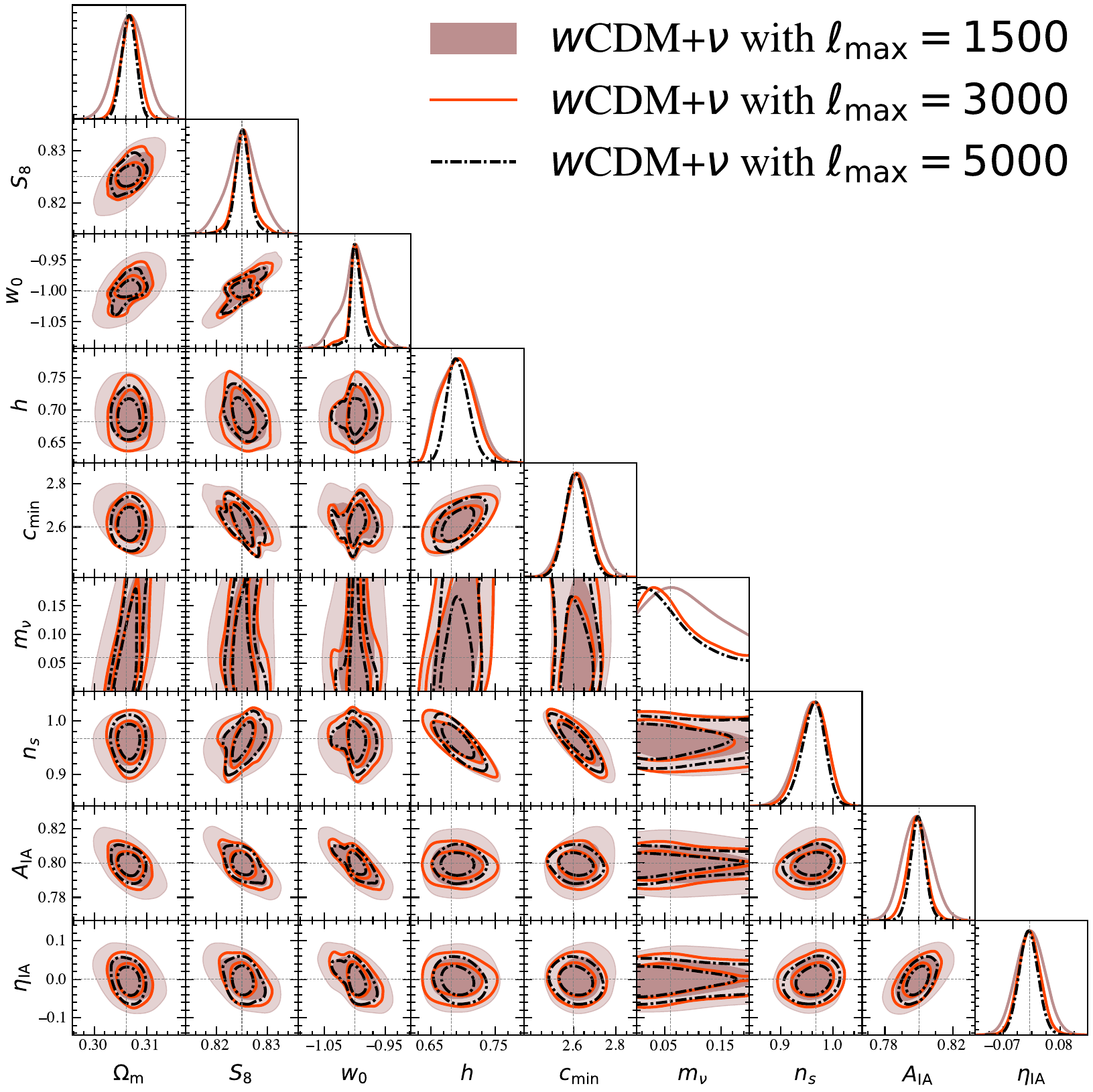}
\caption{Same as in Fig.~\ref{fig:lcdm_comparison}, but for a $w$CDM model.}
\label{fig:wcdm_comparison} 
\end{figure}

\begin{figure}
\centering 
\includegraphics[width=.48\textwidth]{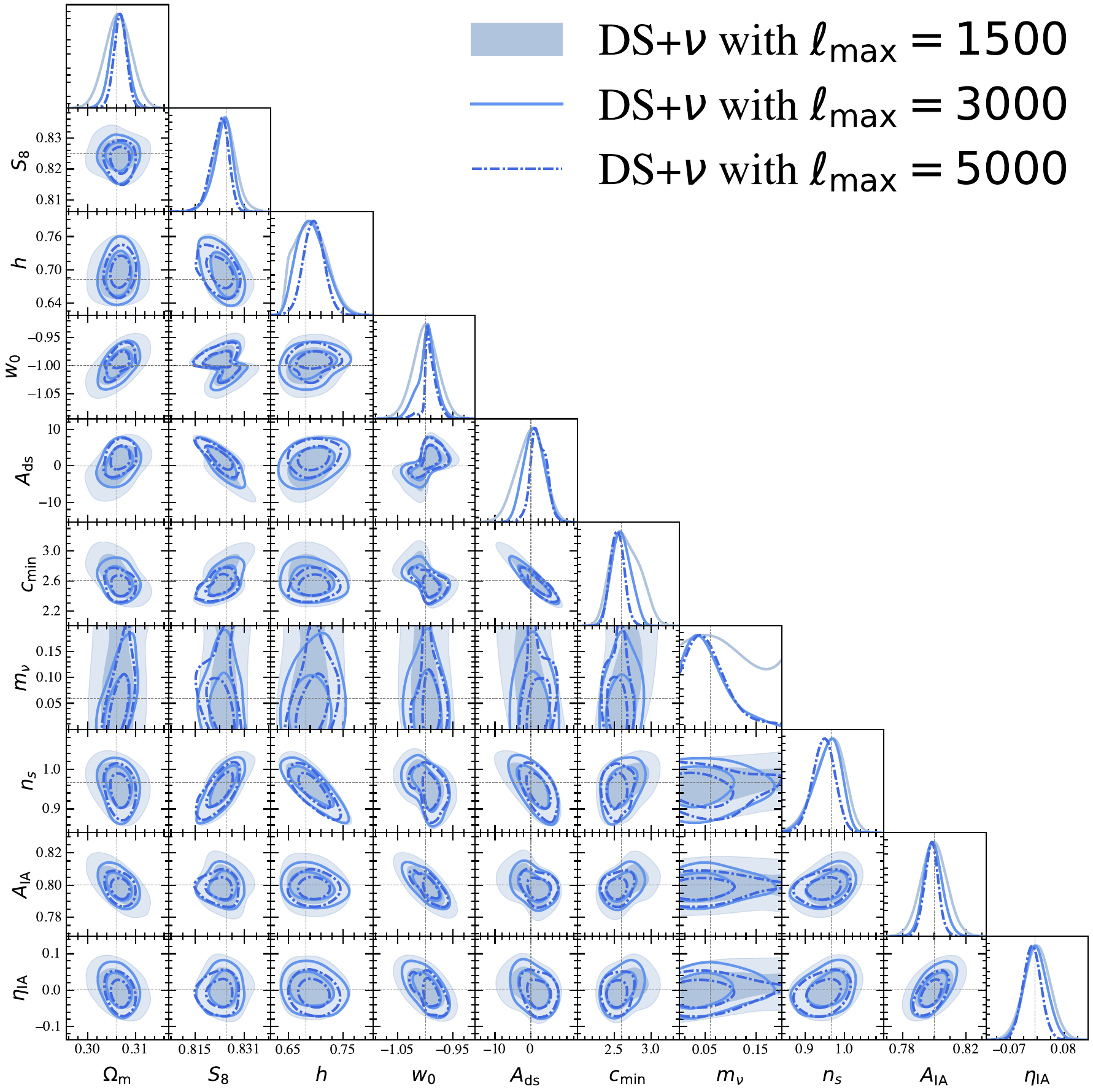}
\caption{Same as in Fig.~\ref{fig:lcdm_comparison}, but for a DS model.}
\label{fig:ds_comparison} 
\end{figure}

\section{Forecasting Stage IV surveys}
\label{sec:forecast}

We consider a survey with the following Stage IV configuration: a sky fraction $f_{\rm sky} = 0.36$, a surface density of galaxies $n_{\rm g}=30$~galaxies/arcmin$^2$ and an observed ellipticity dispersion $\sigma_{\epsilon} = 0.3$. The redshift distributions of galaxies, $n(z)$, are modelled following \citet{1994MNRAS.270..245S}, using $N_{\rm bins} = 10$ tomographic redshift bins  {over a range of $0 \leq z \leq 5$. Each bin is modelled by a Gaussian kernel density estimation with a width of $0.01$.} We compare three cosmological models: $\Lambda$CDM, $w$CDM and DS. 

Regarding the inference setup, our selection of \texttt{NUTS} hyperparameters ensures that each run achieves convergence, controlled by the \texttt{r\_hat} $<1.05$ parameter, in accordance with the Gelman-Rubin criterion \citep{GR_crit1992,2018arXiv181209384V}. We employ the \texttt{chain\_method=`vectorized'} and configure a total of 10 \texttt{num\_chains}, each consisting of at least $10^3$ \texttt{num\_warmup} and \texttt{num\_samples}. For our forecasts, we assume noiseless data vectors, as typical in the literature for forecasts of this type \citep[see e.g.][]{Euclid:2024yrr}. We assume a Gaussian likelihood throughout, with a Gaussian covariance \citep{Joachimi:2007xd}. In future work we will assess the impact of non-Gaussian contributions \citep{Kayo:2012nm,Sato:2013mq,Krause:2016jvl} and super-sample \citep{Li:2014sga, Barreira:2017fjz, lacasa} terms.

\subsection{Fiducial $\Lambda$CDM data}
\label{subsec:F1}

For our first set of forecasts we create a mock data vector described by a $\Lambda$CDM model with massive neutrinos, with parameter values taken from the Planck 2018+BAO \citep{Planck:2018vyg} best-fit analysis (see Table \ref{tab:priors_and_fiducial}). We use the modelling described in subsec.~\ref{subsec:shear_model}; the $\mathcal{C}^{\hat{\gamma} \hat{\gamma}}(\ell)$ spectra of Eq.~(\ref{eq:cell_shear_full}) are binned over $30$-$\ell$ bins between $\ell_{\rm{min}} = 30$ and $\ell_{\rm{max}} = 5000$, and scale cuts are subsequently applied. 

First, we consider a conservative scenario with scales ranging from $\ell_{\rm{min}} = 30$ to $\ell_{\rm{max}} = 1500$. For the second scale cut we extend the upper limit to $\ell_{\rm max} = 3000$. Finally, we also consider an upper limit of $\ell_{\rm max} = 5000$. As a larger $\ell_{\rm max}$ is considered for the scale cut, we expect to achieve greater sensitivity in the signal. 

Our initial model for the analysis is $\Lambda$CDM, including contributions from baryonic feedback and massive neutrinos, i.e.\ the same model used to generate our mock data vector. The priors used in the inference pipeline are listed in Table \ref{tab:priors_and_fiducial}. The input parameters for the emulators in Sec.~\ref{sec:emulators} are assigned uniform distributions, whereas the shifts in the redshift distribution follow a Gaussian prior. The second model we consider is $w$CDM, i.e.\ in practice we employ the same priors used for $\Lambda$CDM, but we also let $w$ free to vary uniformly between $[-1.3, -0.7]$. Finally, we also analyse the $\Lambda$CDM mock data assuming a DS model, with $A_{\rm ds}$ varying uniformly in $[0, 30]$ b/GeV.

Figs.~\ref{fig:lcdm_comparison}, \ref{fig:wcdm_comparison} and \ref{fig:ds_comparison} show results assuming a $\Lambda$CDM, $w$CDM and DS model, respectively. All contour plots in this paper are obtained using \texttt{GetDist} \citep{2019arXiv191013970L}. In each corner plot, filled contours are used for the $\ell_{\mathrm{max}} = 1500$ case, solid contours are used for $\ell_{\mathrm{max}} = 3000$ and dashed contours for $\ell_{\mathrm{max}} = 5000$. Table~\ref{tab:lcdm_fiducial_data} in Appendix \ref{sec:appendix} reports marginalised 1$\sigma$ constraints for key cosmological parameters and for all combinations of scale cuts and cosmological models. We do not include the final constraints on the redshift distribution shifts $D_{z_{i}}$ in the figures and table, as we find these are prior-dominated and do not show strong degeneracies with any other cosmological or nuisance parameter.

For all three models, we notice that increasing the scale cut from $\ell_{\rm max} = 1500$ to $\ell_{\rm max} = 3000$ leads to tighter constraints on $\Omega_{\rm m}$ and $S_8$, as expected, as these are the parameters best constrained by cosmic shear. However, going from $\ell_{\rm max} = 3000$ to $\ell_{\rm max} = 5000$ does not lead to major improvements in the constraints. Similarly, constraints on $m_{\nu}$ and $w_0$ reduce by a factor 2 going from $\ell_{\mathrm{max}} = 1500$ to $3000$, but the inclusion of multipoles up to $\ell_{\mathrm{max}} = 5000$ does not lead to significant improvements. This is in line with the results of \citet{SpurioMancini:2023mpt}, who found that including these highly nonlinear scales only mildly improves the constraining power on cosmological parameters if baryonic feedback parameters are \textit{a priori} unconstrained (as is the case here for $c_{\rm min}$), as baryonic parameters tend to absorb a large fraction of the constraining power. This motivates further research into ways to constrain the priors on these parameters \citep[see e.g.][for recent examples]{Arico:2024pvt,Bigwood,PhysRevLett.133.051001}.
 
 {When assuming a DS model, we observe that for all scale cuts, $\ell_{\rm max}$, the mean value is $w_0 > -1$, leading to $A_{\rm ds} > 0$, while the 1D marginalised posterior remains consistent with the fiducial value. We attribute the observed shift in $|A_{\rm ds}|$ primarily to parameter degeneracies, which may introduce biases, especially with baryonic feedback ($ c_{\rm min}$), as shown in \autoref{fig:ds_comparison}}. It is important to highlight the characteristic 'butterfly'-shaped contours in the $w_0$-$A_{\text{ds}}$ plane associated with this model \citep{Tsedrik:2022cri,Carrilho:2022mon,Carrion:2024itc} due to the degeneracy between these two parameters -- cf. Eq.~(\ref{eq:Interaction_term_eff}). Overall, we find that $A_{\rm ds}$ is constrained with a marginalised $1\sigma$ uncertainty that is roughly an order of magnitude smaller than that obtained in our previous cosmic shear analysis of KiDS data \citep{Carrion:2024itc}, highlighting the promising constraining power of Stage IV configurations.

Interestingly, for all scale cuts the DS model produces uncertainties on the baryon feedback parameter $c_{\min}$ larger than both $\Lambda$CDM and $w$CDM. This can be attributed to the known degeneracy between the effects of baryonic feedback and the DS interaction on the matter power spectrum (see Fig.~\ref{fig:spectra_impact}). A similar degeneracy between the dark sector interaction and the effect of massive neutrinos on the matter power spectrum explains why the constraints lead to a well-defined 1D $m_\nu$ posterior distribution only for scale cuts at $\ell_{\mathrm{max}} = 3000$ and $5000$. Both of these degeneracies were already highlighted in \cite{Carrilho:2021rqo}, and we confirm their impact on contours here.

As expected, given the $\Lambda$CDM mock data vector considered in this section, our results indicate a preference for the $\Lambda$CDM model; this is evident in the values of both the contours and the log Bayes factors $\log B$, the latter reported in Table \ref{tab:harmonic_values}. For a given scale cut, $\Lambda$CDM generally shows tighter constraints compared to the other two models, particularly in $\Omega_{\rm m}$ and $S_8$, which is expected given the fewer degrees of freedom in $\Lambda$CDM, as compared to $w$CDM and DS. The values of the log Bayes factors produced by \texttt{harmonic} present a coherent picture, confirming $w$CDM and DS to be disfavoured compared to $\Lambda$CDM. As expected, the DS model is even more disfavoured than the $w$CDM model due to the introduction of an additional parameter. We also note a trend in the log Bayes factor computed between two models, namely that this value increases as the $\ell_{\rm max}$ increases, for each pair of models compared in the analysis. This is also expected, since more information becomes available as we include more non-linear scales. We note, however, that the uncertainty on the log Bayes factor also increases with $\ell_{\rm max}$. This can be attributed to the shape of the posteriors shrinking at higher values of $\ell_{\rm max}$, which makes it more challenging for the normalising flow employed by \texttt{harmonic} to have tails narrower than its target distribution, i.e.\ the posterior.

We stress here that the reason for the particularly high values of the log Bayes factors reported in this work is due to the noiseless data vectors considered for our forecasts, as customary in the literature \citep[see e.g.][]{Euclid:2024yrr}. In realistic applications, noisy data would lead to smaller values of the log Bayes factor. However, even with noisy mock data we expect to see a similar trend in the impact of non-linear scales on the values of the log Bayes factor. This demonstrates the importance of including non-linear information in cosmological inference pipelines not only to improve parameter estimation, but also model comparison.

\subsection{Fiducial DS data}
\label{subsec:F2}

The second set of forecasts we consider assumes a mock dataset generated under the assumption of a DS fiducial model; specifically, we assume $A_{\rm ds} = 10.6$ b/GeV for our mock data vector (see Table \ref{tab:priors_and_fiducial}). We choose this value as this is the best fit from our previous analysis \citep{Carrion:2024itc}.

\begin{figure}
\centering 
\includegraphics[width=.48\textwidth]{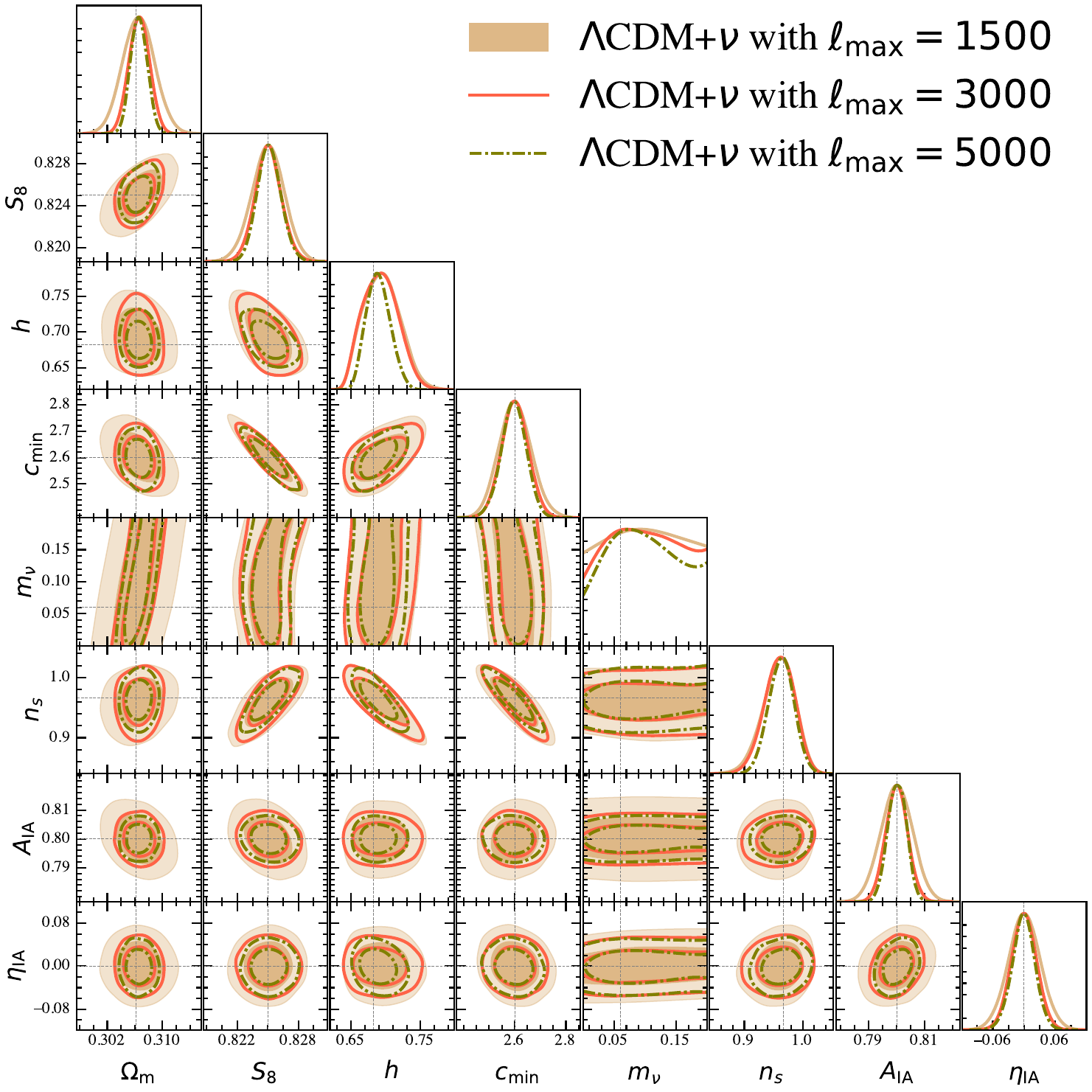}
\caption{68\% and 95\% 2D and 1D marginalised posterior contours for key cosmological and nuisance parameters for the $\Lambda$CDM model using three different scale-cuts. The thin dashed-grey straight lines indicate the DS fiducial values used to generate the mock data vector.}
\label{fig:lcdm_comparison_f2} 
\end{figure}

\begin{figure}
\centering 
\includegraphics[width=.48\textwidth]{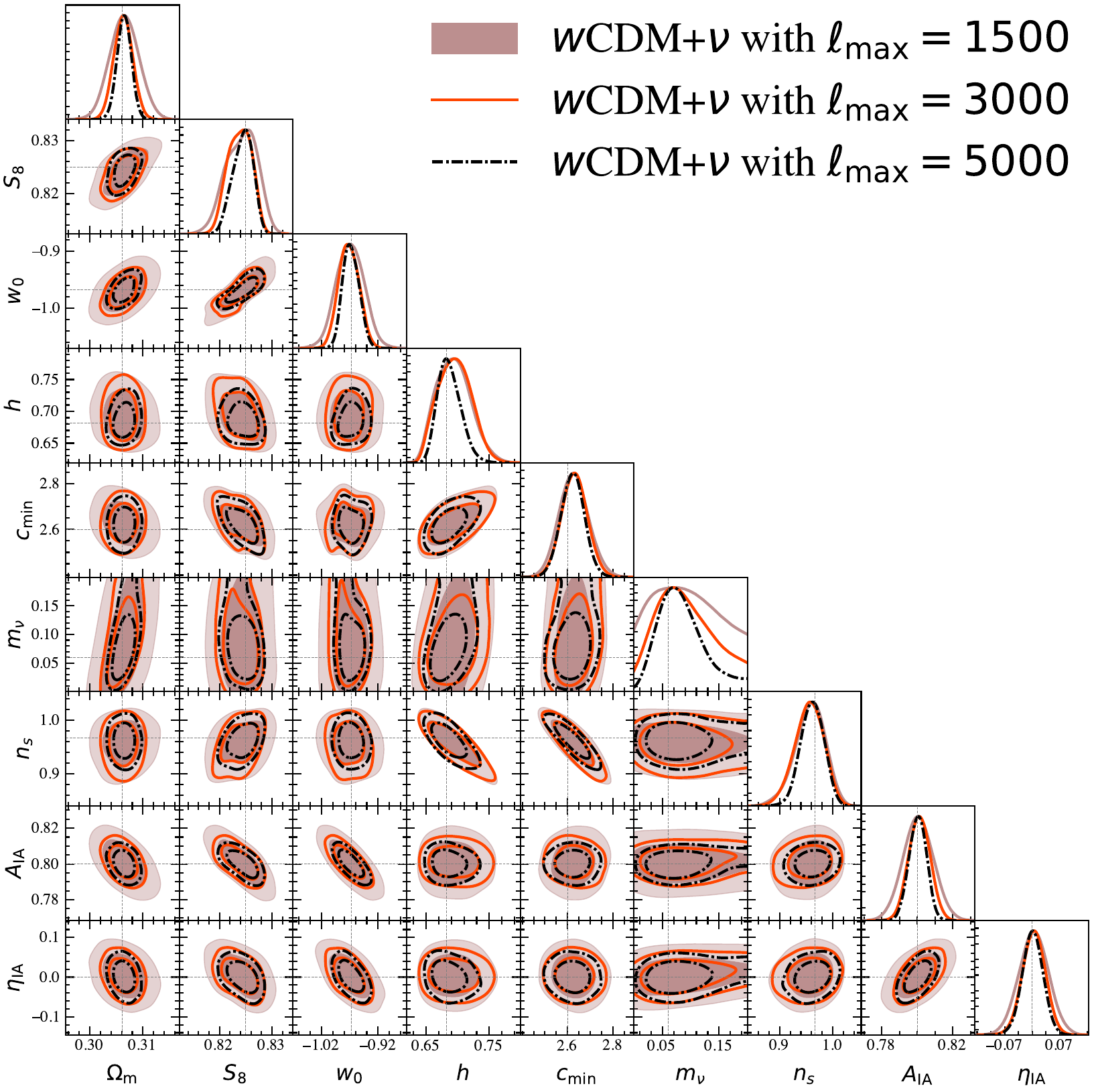}
\caption{Same as in Fig.~\ref{fig:lcdm_comparison_f2}, but for a $w$CDM model.}
\label{fig:wcdm_comparison_f2} 
\end{figure}

\begin{figure}
\centering 
\includegraphics[width=.48\textwidth]{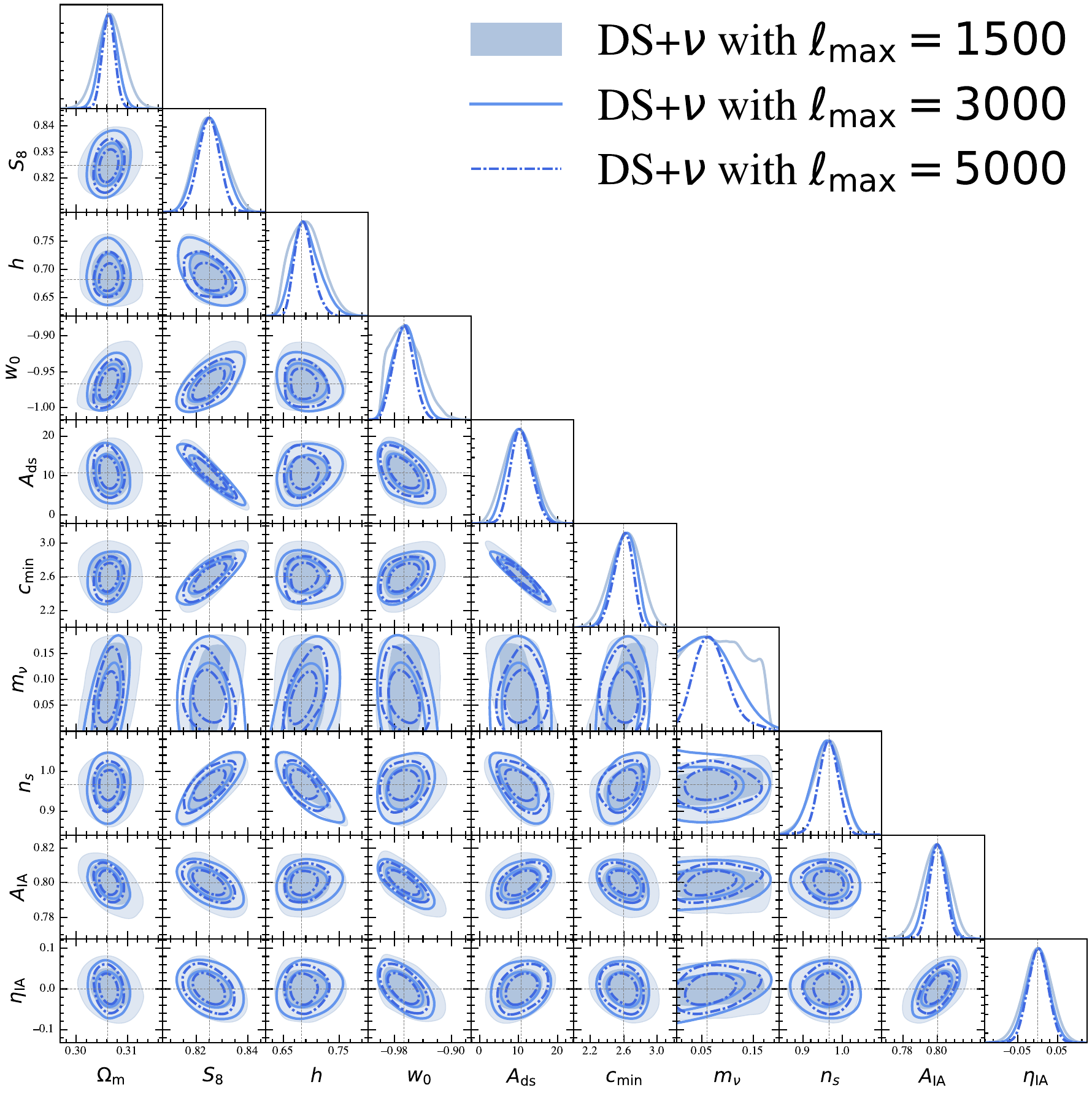}
\caption{Same as in Fig.~\ref{fig:lcdm_comparison_f2}, but for a DS model.}
\label{fig:ds_comparison_f2} 
\end{figure}

The results assuming a $\Lambda$CDM model are shown in Fig.~\ref{fig:lcdm_comparison_f2}, which indicates the best-fit values of the posteriors closely match the fiducial values. For the $w$CDM model, presented in Fig.~\ref{fig:wcdm_comparison_f2}, we note a very moderate bias in the baryonic feedback parameter $c_{\rm min}$, which we attribute to the degeneracy between baryonic feedback and DS interaction mentioned in Sec.~\ref{subsec:F1}. Moreover, in the $w$CDM case, the fiducial value of $w_0$ in the mock data imposes a better constraint than our first analysis and shows a correlation between $w_0$ and $S_8$. Finally, Fig.~\ref{fig:ds_comparison_f2} shows the posterior contours obtained modelling our theory predictions with a DS model, with which we generated the mock data vector. The constraints on the DS amplitude parameter $A_{\rm ds}$ improve slightly when using larger values of the maximum angular scale: $A_{\rm ds}$ is constrained with a 1$\sigma$ relative uncertainty $\sigma_{A_{\rm ds}}/A_{\rm ds} \sim 36\%, 30\%, 24\%$ for $\ell_{\rm max}=1500, 3000, 5000$, respectively.

Importantly, the typical “butterfly” structure is absent in this case, as the parameter $w$ is constrained to values $w>-1$. Additionally, an anti-correlation is observed between the parameter $A_{\rm ds}$ and both $S_8$ and $c_{\rm min}$. Table~\ref{tab:ds_fiducial_data} sums up the marginalised 1$\sigma$ constraints for all parameters, reflecting the three combinations of scale cuts over all cosmological models.
Regarding the model comparison results, the values of the log Bayes factor reported in Table ~\ref{tab:harmonic_values} confirm also for this set of forecasts the same trend reported in the Sec.~\ref{subsec:F2}: DS is correctly identified by \texttt{harmonic} as the better fit to the data compared to both $w$CDM and $\Lambda$CDM, and values of the log Bayes factors increase as one includes more non-linear scales in the analysis.

\begin{table}
\centering
\caption{Evidence values for different scale-cuts $\ell_{\rm max}$, for $\Lambda$CDM and DS mock data. We stress that, as typical in the literature for forecasts of this type, the fiducial data vector is assumed noiseless, which leads to high values of the log Bayes factor (see text for details).}
\label{tab:harmonic_values}
\renewcommand{\arraystretch}{1.75}
\setlength{\tabcolsep}{3.5pt}
\begin{tabular}{c||c c||c c}
\multicolumn{1}{c}{~} & \multicolumn{2}{c}{\textbf{$\Lambda$CDM mock data}} & \multicolumn{2}{c}{\textbf{DS mock data}} \\ 
$\ell_{\rm max}$ & $ \log \left( \dfrac{\mathcal{Z}_{\Lambda \rm CDM}}{\mathcal{Z}_{w\rm CDM}} \right)$ & $ \log \left( \dfrac{\mathcal{Z}_{\Lambda \rm CDM}}{\mathcal{Z}_{\rm DS}} \right)$ & $ \log \left( \dfrac{\mathcal{Z}_{\rm DS}}{\mathcal{Z}_{\rm \Lambda CDM}} \right)$ & $ \log \left( \dfrac{\mathcal{Z}_{\rm DS}}{\mathcal{Z}_{w \rm CDM}} \right)$\\ 
\hline \hline
$1500$ & $2.39^{\, +0.13}_{-0.14}$ & $4.22^{\, +0.41}_{-0.45}$ 
& $3.39^{\, +0.53}_{-0.62}$ & $1.40^{\, +0.34}_{-0.41}$ \\
$3000$ & $3.23^{\, +0.84}_{-1.13}$ & $5.30^{\, +0.45}_{-0.48}$ 
& $5.64^{\, +0.41}_{-0.35} $ & $4.20^{\, +0.32}_{-0.28}$  \\ 
$5000$ & $4.04^{\, +1.48}_{-1.18}$ & $
5.89^{\, +1.42}_{-1.86}$ 
& $6.10^{\, +1.15}_{-1.34}$ & $4.57^{\,+ 0.72}_{-0.81}$ \\ 
\end{tabular}
\end{table}

\section{Conclusion and outlook}
\label{sec:conclusion}

In this paper, we produced cosmic shear forecasts for a Stage IV survey configuration using a fast and accurate emulator of the dark scattering (DS) non-linear matter power spectrum obtained with \texttt{CosmoPower}, incorporating the impact of baryonic feedback and massive neutrinos into the halo model reaction. By embedding this emulator within a fully-differentiable inference pipeline that can be run on GPUs, we drastically accelerate Bayesian inference via gradient-based sampling. Such enhancements are crucial for extracting precise and accurate insights from the extensive datasets anticipated from Stage IV galaxy surveys; moreover, the accelerated analysis allows us to consider different fiducial models, thereby exploring the different possibilities that the data may present.

Our forecasts demonstrate that several key parameters of the model are moderately sensitive to different $\ell_{\rm max}$, showing the impact of scale cuts in the final cosmological constraints. For example, considering a DS model for both the mock data and the modelling in the inference pipeline, we find that the relative 1$\sigma$ uncertainty on the DS amplitude parameter $A_{\rm ds}$ goes from 36\% to 24\% as we increase $\ell_{\rm max}$ from 1500 to 5000. However, we also highlighted the importance of setting strong priors on baryonic feedback parameters \cite[e.g.][]{Arico:2024pvt, Bigwood, PhysRevLett.133.051001}, to ensure that including more non-linear scales in the analysis effectively leads to a significant increase in constraining power.

An important difference with respect to our previous work presented in \citet{Carrion:2024itc} is the inclusion of varying massive neutrinos in the inference pipeline. Additionally, we perform model comparison using the learnt harmonic mean estimator, allowing us to straightforwardly estimate the evidence for the DS, $w$CDM and $\Lambda$CDM models through a process completely decoupled from the parameter estimation pipeline. It is important to emphasise that our \texttt{JAX}-based pipeline can be easily adapted to study other cosmological models beyond $\Lambda$CDM. Furthermore, the pipeline can scale to even higher dimensional parameter spaces \citep{Piras:2024dml}, making it suitable for more complex cosmological analyses involving additional systematic parameters or new physics. 

The constraints presented in this forecast show a promising future for constraining DS models with Stage IV cosmic shear data. The forecasts show an increase in constraining power on $A_{\rm ds}$ of approximately an order of magnitude over Stage III cosmic shear configurations \citep{Carrion:2024itc}. As a next step, we will combine cosmic shear with galaxy clustering in a 3x2pt analysis \citep[e.g.][]{Heymans_2021, Abbott_2022} to forecast the further improvement expected from a joint analysis of these probes. In addition, we will include a more complete modelling of baryonic feedback, e.g. using the \texttt{BCEmu} \citep{Giri:2021qin} or \texttt{Bacco} \citep{Arico_2021} emulators.

\section*{Acknowledgements}

 {We want to thank the anonymous referee for several valuable comments.} We are grateful to Alkistis Portsidou and Jason McEwen for insightful feedback on this work. We acknowledge access to Cuillin, a computing cluster of the Royal Observatory, University of Edinburgh. KC acknowledges support from a CONAHCyT studentship and from grant No. CBF2023-2024-162. KC is grateful to Royal Holloway, University of London for hospitality. DP was supported by the SNF Sinergia grant CRSII5-193826 “AstroSignals: A New Window on the Universe, with the New Generation of Large Radio-Astronomy Facilities”. KC and JCH acknowledge support from grant CONAHCyT CBF2023-
2024-162. JCH acknowledges support from program UNAM-PAPIIT grant IG102123 ``Laboratorio de Modelos y Datos (LAMOD) para proyectos de Investigación Científica: Censos Astrofísicos".
\section*{Data Availability}

The trained emulators are available at \href{https://github.com/karimpsi22/DS-emulators}{\texttt{DS-emulators}}.

\bibliographystyle{mnras}
\bibliography{mybib} 



\appendix

\section{Appendix A}
\label{sec:appendix}
We report mean and 68\% credible intervals for the two fiducial cases considered in this work in Table \ref{tab:lcdm_fiducial_data} and Table \ref{tab:ds_fiducial_data}.

\begin{table*}
    \centering
    \caption{Mean and 68\% marginalised credible intervals for key parameters characterising the $\Lambda$CDM, $w$CDM and DS models, for scale cuts $\ell_{\rm max} = 1500, 3000$, and 5000. The mock data vector is generated assuming a $\Lambda$CDM fiducial.}
\label{tab:lcdm_fiducial_data}
    \renewcommand{\arraystretch}{1.75}
    \setlength{\tabcolsep}{3.5pt}
    \begin{adjustbox}{max width=\textwidth}
    \begin{tabular}{|c|c|c|c||c|c|c||c|c|c|}
        \multicolumn{10}{c}{{\Large\textbf{$\Lambda$CDM mock data}}} \\ \hline
        \multirow{2}{*}{\textbf{Parameter}} & \multicolumn{3}{c||}{\textbf{$\ell_{\rm max} = 1500$}} & \multicolumn{3}{c||}{\textbf{$\ell_{\rm max} = 3000$}} & \multicolumn{3}{c|}{\textbf{$\ell_{\rm max} = 5000$}} \\ 
        \cline{2-10} \cline{2-10}                                    
        & \textbf{$\Lambda$CDM} & \textbf{$w$CDM} & \textbf{DS} & \textbf{$\Lambda$CDM} & \textbf{$w$CDM} & \textbf{DS} & \textbf{$\Lambda$CDM} & \textbf{$w$CDM} & \textbf{DS} \\ \hline
        
        $\omega_{\rm b}$  
        & $0.023 \pm 0.002$ & $0.023 \pm 0.002$ & $0.023 \pm 0.002$ 
        & $0.023 \pm 0.002$ & $0.023 \pm 0.002$&  $0.023 \pm 0.002$ 
        & $0.022 \pm 0.002$ & $0.023 \pm 0.002$ & $0.023 \pm 0.002$ \\ \hline
        
        $\omega_{\rm cdm}$  
        & $0.124 \pm 0.010$ & $0.124 \pm  0.010$ & $0.123 \pm 0.010$ 
        & $0.123 \pm 0.009$ & $0.123 \pm 0.009$ & $0.124 \pm 0.010$ 
        & $0.122 \pm 0.006$ & $0.123 \pm 0.006$ & $0.125 \pm 0.007$ \\ \hline
        
        $h$  
        & $0.69 \pm 0.03$ & $0.69 \pm 0.03$ & $0.69 \pm 0.03$ 
        & $0.69 \pm 0.03$ & $0.69 \pm 0.025$ & $0.69 \pm 0.03$ 
        & $0.69 \pm 0.02$ & $0.69 \pm 0.02$ & $0.70 \pm 0.02$ \\ \hline
        
        $n_s$  
        & $0.96 \pm 0.03$ & $0.96 \pm 0.03$ & $0.96 \pm 0.04$ 
        & $0.96 \pm 0.03$ & $0.96 \pm 0.03$ & $0.96 \pm 0.03$ 
        & $0.96 \pm 0.02$ & $0.96 \pm 0.02$ & $0.95 \pm 0.03$ \\ \hline

        $\Omega_{\rm m}$  
        & $0.306 \pm 0.002$ & $0.306 \pm 0.002$ &  $0.306 \pm 0.003$
        & $0.306 \pm 0.001$ & $0.306 \pm 0.001$ &  $0.306 \pm 0.001$
        & $0.306 \pm 0.001$ & $0.306 \pm 0.001$ & $0.306 \pm 0.001$ \\ \hline

        $S_8$  
        & $0.825 \pm 0.002$ & $0.825 \pm 0.003$ & $0.824 \pm 0.003$ 
        & $0.825 \pm 0.001$ & $0.825 \pm 0.002$ & $0.824 \pm 0.003$ 
        & $0.825 \pm  0.001$ & $0.825 \pm 0.002$ & $0.823 \pm 0.003$ \\ \hline
        
        $m_\nu$  
        & $0.10 \pm 0.06$ & $0.09 \pm 0.05$ & $0.09 \pm 0.06$ 
        & $0.10 \pm 0.06$ & $0.08 \pm 0.05$ & $0.06 \pm 0.04$ 
        & $0.10 \pm 0.06$ & $0.08 \pm 0.05$ & $0.06 \pm 0.04$ \\ \hline
        
        $w_0$  
        & $-$ & $-0.997 \pm 0.025$ &  {$-0.999 \pm 0.022$} 
        & $-$ & $-0.996 \pm 0.015$ & $-0.997 \pm 0.017$ 
        & $-$ & $-0.997 \pm 0.012$ & $-0.991 \pm 0.012$ \\ \hline
        
        $|A_{\rm ds}|$  
        & $-$ & $-$ &  {$3.02 \pm 2.25$} 
        & $-$ & $-$ & $2.33 \pm 1.73$ 
        & $-$ & $-$ & $2.47 \pm 1.81$ \\ \hline
        
        $c_{\rm min}$  
        & $2.60 \pm 0.06$ & $2.63 \pm 0.07$ & $2.65 \pm 0.18$ 
        & $2.60 \pm 0.05$ & $2.62 \pm 0.06$ & $2.60 \pm 0.13$ 
        & $2.60 \pm 0.05$ & $2.61 \pm 0.05$ & $2.54 \pm 0.09$ \\ \hline
        
        
        \hline
    \end{tabular}
    \end{adjustbox}
\end{table*}

\begin{table*}
    \centering
    \caption{Same as Table~\ref{tab:lcdm_fiducial_data}, but for a DS fiducial.}
    \label{tab:ds_fiducial_data}
    \renewcommand{\arraystretch}{1.75}
    \setlength{\tabcolsep}{3.5pt}
    \begin{adjustbox}{max width=\textwidth}
    \begin{tabular}{|c|c|c|c||c|c|c||c|c|c|}
        \multicolumn{10}{c}{{\Large\textbf{DS mock data}}} \\ \hline
        \multirow{2}{*}{\textbf{Parameter}} & \multicolumn{3}{c||}{\textbf{$\ell_{\rm max} = 1500$}} & \multicolumn{3}{c||}{\textbf{$\ell_{\rm max} = 3000$}} & \multicolumn{3}{c|}{\textbf{$\ell_{\rm max} = 5000$}} \\ 
        \cline{2-10} \cline{2-10}                                    
        & \textbf{$\Lambda$CDM} & \textbf{$w$CDM} & \textbf{DS} & \textbf{$\Lambda$CDM} & \textbf{$w$CDM} & \textbf{DS} & \textbf{$\Lambda$CDM} & \textbf{$w$CDM} & \textbf{DS} \\ \hline
        
        $\omega_{\rm b}$  
        & $0.023 \pm 0.002$ & $0.023 \pm 0.002$ & $0.023 \pm 0.002$
        & $0.023 \pm 0.002$ & $0.023 \pm 0.002$ & $0.023 \pm 0.002$
        & $0.023 \pm 0.002$ & $0.022 \pm 0.002$ & $0.023 \pm 0.002$ \\ \hline
        
        $\omega_{\rm cdm}$  
        & $0.124 \pm 0.009$ & $0.124 \pm 0.010$ & $0.123 \pm 0.011$ 
        & $0.123 \pm 0.009$ & $0.124 \pm 0.010$ & $0.123 \pm 0.009$
        & $0.122 \pm 0.006$ & $0.121 \pm 0.007$ & $0.121\pm 0.006$ \\ \hline 
        
        $h$  
        & $0.69 \pm 0.03$ & $0.69 \pm 0.03$ & $0.69 \pm 0.03$
        & $0.69 \pm 0.03$ & $0.70 \pm 0.03$ & $0.69 \pm 0.02$
        & $0.69 \pm 0.02$ & $0.69 \pm 0.02$ & $0.69 \pm 0.02$ \\ \hline
        
        $n_s$  
        & $0.96 \pm 0.03$ & $0.96 \pm 0.03$ & $0.96 \pm 0.04$
        & $0.96 \pm 0.03$ & $0.96 \pm 0.03$ & $0.96 \pm 0.04$
        & $0.96 \pm 0.02$ & $0.96 \pm 0.02$ & $0.96 \pm 0.03$ \\ \hline

        $\Omega_{\rm m}$  
        & $0.306 \pm 0.002$ & $0.306 \pm 0.002$ &  $0.306 \pm 0.002$
        & $0.306 \pm 0.001$ & $0.306 \pm 0.001$ &  $0.306 \pm 0.001$
        & $0.306 \pm 0.001$ & $0.306 \pm 0.001$ & $0.306 \pm 0.001$ \\ \hline
        
        $S_8$  
        & $0.825 \pm 0.002$ & $0.824 \pm 0.003$ & $0.826 \pm 0.006$
        & $0.825 \pm 0.001$ & $0.824 \pm 0.002$ & $0.825 \pm 0.005$
        & $0.825 \pm 0.001$ & $0.824 \pm 0.002$ & $0.825 \pm 0.004$ \\ \hline
        
        $m_\nu$  
        & $0.10 \pm 0.06$ & $0.09 \pm 0.05$ & $0.09 \pm 0.05$
        & $0.10 \pm 0.06$ & $0.09 \pm 0.05$ & $0.07 \pm 0.04$
        & $0.10 \pm 0.05$ & $0.08 \pm 0.04$ & $0.07 \pm 0.03$ \\ \hline
        
        $w_0$  
        & $-$ & $-0.969 \pm 0.025$ & $-0.963 \pm 0.022$
        & $-$ & $-0.972 \pm 0.018$ & $-0.965 \pm 0.017$
        & $-$ & $-0.968 \pm 0.014$ & $ -0.967 \pm 0.014$ \\ \hline
        
        $|A_{\rm ds}|$  
        & $-$ & $-$ & $10.20  \pm 3.77$
        & $-$ & $-$ & $10.36 \pm 3.16$
        & $-$ & $-$ & $10.74 \pm 2.67$ \\ \hline
        
        $c_{\rm min}$  
        & $2.60 \pm 0.06$ & $2.63 \pm 0.06$ & $2.62 \pm 0.18$
        & $2.60 \pm 0.05$ & $2.63 \pm 0.06$ & $2.61 \pm 0.13$
        & $2.60 \pm 0.05$ & $2.62 \pm 0.05$ & $2.59 \pm 0.11$ \\ \hline
        
        
        \hline

    \end{tabular}
    \end{adjustbox}
\end{table*}

\bsp	
\label{lastpage}
\end{document}